\documentclass[aps,prb,twocolumn]{revtex4-1}%
\usepackage{lineno,xcolor}
\usepackage{graphicx}
\usepackage{bm}
\usepackage[breaklinks=true]{hyperref}
\usepackage{enumitem}
\usepackage{epstopdf}
\usepackage{amsmath}
\usepackage{amsfonts}
\usepackage{amssymb}%
\providecommand{\U}[1]{\protect\rule{.1in}{.1in}}
\hypersetup{bookmarksnumbered, pdfpagemode=UseOutlines,
pdfauthor={A.\ Aqeel},
pdftitle={Spin-Hall magnetoresistance and spin Seebeck effect in a chiral magnetic insulator}, pdfdisplaydoctitle,
colorlinks=true, citecolor=blue, filecolor=blue, linkcolor=blue, urlcolor=blue}

\usepackage{hyperref}
\usepackage[bottom]{footmisc}

\begin{document}


\date{\today}
\title{Electrical detection of spiral spin structures in Pt$|$Cu$_2$OSeO$_3$ heterostructures}
\author{A.\ \surname{Aqeel}}
\affiliation{Zernike Institute for Advanced Materials, University of Groningen, Nijenborgh
4, 9747 AG Groningen, The Netherlands}
\author{N.\ \surname{Vlietstra}}
\affiliation{Zernike Institute for Advanced Materials, University of Groningen, Nijenborgh
4, 9747 AG Groningen, The Netherlands}
\author{A.\ \surname{Roy}}
\affiliation{Zernike Institute for Advanced Materials, University of Groningen, Nijenborgh
4, 9747 AG Groningen, The Netherlands}
\author{M.\ \surname{Mostovoy}}
\affiliation{Zernike Institute for Advanced Materials, University of Groningen, Nijenborgh
4, 9747 AG Groningen, The Netherlands}
\author{B.\ J.\ \surname{van Wees}}
\affiliation{Zernike Institute for Advanced Materials, University of Groningen, Nijenborgh
4, 9747 AG Groningen, The Netherlands}
\author{T.\ T.\ M.\ \surname{Palstra}}
	\email[e-mail: ]{t.t.m.palstra@rug.nl}
\affiliation{Zernike Institute for Advanced Materials, University of Groningen, Nijenborgh
4, 9747 AG Groningen, The Netherlands}

\maketitle

\textbf{The interaction between the itinerant spins in metals and localized spins in magnetic insulators thus far has only been explored in collinear spin-systems, such as garnets. Here, we report the spin-Hall magnetoresistance (SMR)~\cite{Althammer2013,Vlietstra2013,Nakayama2013,Isasa2014} sensitive to the surface  magnetization~\cite{Isasa2014,Chen2013,JiaSTT2011} of the spin-spiral material, Cu$_2$OSeO$_3$. We experimentally  demonstrate that the angular  dependence of the SMR changes drastically at the transition between the helical spiral and the conical spiral phases. Furthermore, the sign and magnitude of the SMR in the conical spiral state are  controlled by the cone angle. We show that this complex behaviour can be qualitatively explained within the SMR theory  initially developed for collinear magnets. In addition, we studied the spin Seebeck effect (SSE)~\cite{Uchida_nmat_2010,Schreier2013,Aqeel2014}, which is sensitive to the bulk magnetization~\cite{Agrawal2014,Kehlberger2015,Rezende2016}. It originates from the conversion of thermally excited low-energy spin waves in the magnet, known as magnons, into the spin current in the adjacent metal contact (Pt). The SSE displays unconventional behavior where not only the magnitude but also the phase of the SSE vary with the applied magnetic field. }

\begin{figure*}[Hptb]
\centering
\includegraphics[width=0.99\textwidth, clip]{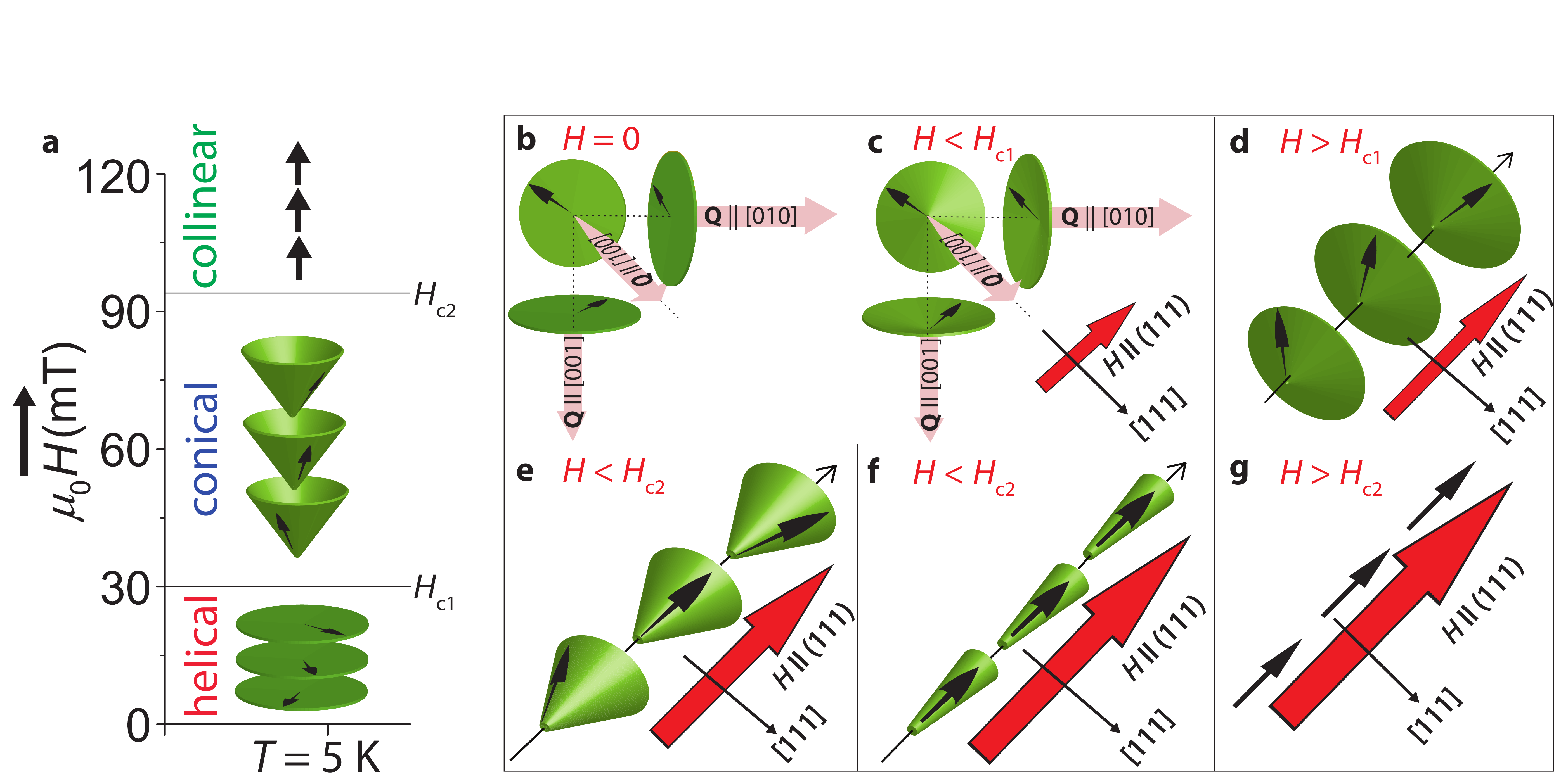}\caption{\textbf{Phase diagram and magnetic spin structures in CSO}. \textbf{a}, A schematic illustration of the magnetic phase diagram of CSO as a function of applied magnetic field at 5~K. Here $H_{c1}$ and $H_{c2}$ represent the fields at which the magnetic transitions occur.  At $H_{c1}$, the helical state with three magnetic domains converts to a single domain conical state oriented along $\bm H$. At $H_{c2}$, conical to collinear ferrimagnetic transition occurs. \textbf{b-g}, The transformation from helical (\textbf{b,c}) to conical (\textbf{d-f}) and then to the FM state (\textbf{g}) under an  increasing applied magnetic field. Here, round brackets ( ) and square brackets [ ] indicate the planes and directions of the unit cell, respectively.}%
\label{fig:1}%
\end{figure*}

Cu$_2$OSeO$_3$ (CSO) belongs to the family of cubic chiral magnets~\cite{Bos2008}, including MnSi, Fe$_{0.5}$Co$_{0.5}$Si and FeGe, in which magnetic skyrmions have been found recently~\cite{Nagaosa2013}.  Unlike the itinerant magnet MnSi, CSO is a ferrimagnetic Mott insulator with a  band gap of 2.1~eV ~\cite{Janson2014}. Yet the magnetic phase diagram of CSO, showing a variety of non-collinear spin states  (see Fig.~\ref{fig:1}a),  is  similar to those of other cubic chiral magnets. At low applied magnetic fields CSO is in an incommensurate helical spiral state  with a long period of $\sim$50~nm~\cite{Seki2012} stabilized by the relativistic Dzyaloshinskii-Moriya interaction~\cite{Bak1980}. In this multi-domain state the spiral wave vector $\bm Q$ can be oriented along any of the three equivalent directions: [100], [010] or [001]~\cite{Adams2012}. Above the critical field, $H_{c1}$, a transition into a conical spiral state occurs, in which $\bm Q \parallel \bm H$. As the field increases, the cone angle $\theta$ becomes smaller. Eventually, $\theta$ becomes 0 at the second critical field, $H_{c2}$, which marks the transition to the field-induced collinear ferrimagnetic (FM) state. The evolution of the magnetic structure of CSO under an increasing magnetic field is depicted in Figs.~\ref{fig:1}b-g.  

\begin{figure*}[Bpht]
\includegraphics[width=0.99\textwidth,natwidth=310,natheight=342]{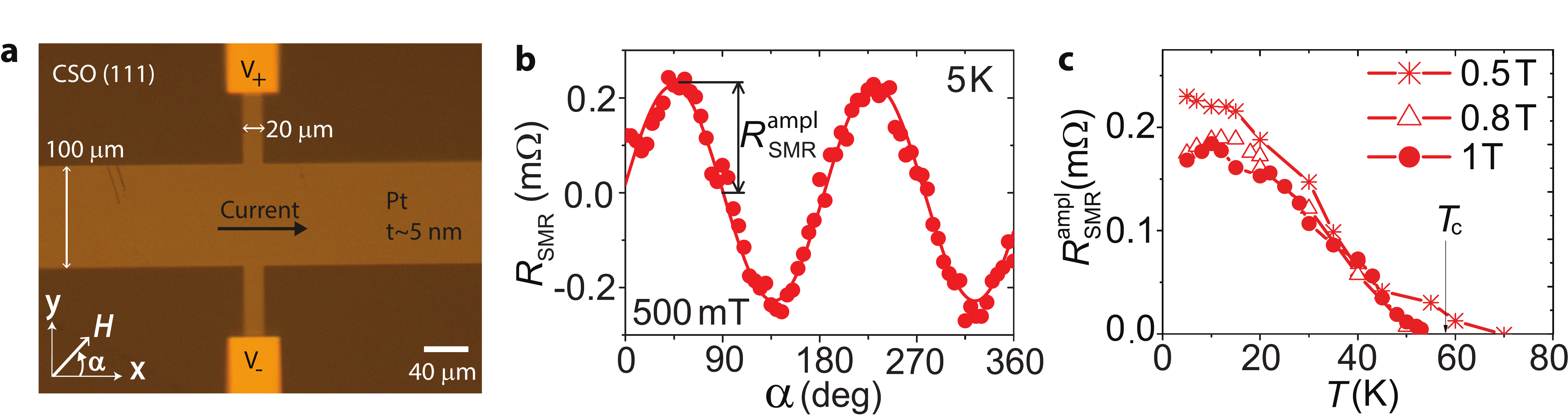}\caption{ \textbf{Detection of the spin-Hall magnetoresistance (SMR)}. \textbf{a}, Optical image showing the Hall-bar structure with 5~nm thick Pt deposited on a CSO crystal. \textbf{b},  Angular dependence of the SMR signal,   $R_{\rm SMR}$, measured by contacting the Hall-bar structure as marked in \textbf{a}. The solid line shows a $\sin(2\alpha)$ fit. In this plot the signals caused by the ordinary Hall effect have been subtracted. \textbf{c}, Temperature dependence of the SMR signal $R_{\rm SMR}^{\rm ampl}$ at several magnetic field strengths. Data are acquired for device S1.  }%
\label{fig:2}%
\end{figure*}

Recently,  the coupling of magnetization to spin, charge and heat currents was much studied in the context of next generation spintronic applications.  In this development, the spin-Hall magnetoresistance (SMR) and spin Seebeck effect (SSE) are very prominent, as they provide information on the magnetization of an insulating magnetic layer by purely electrical measurements.  The underlying physics of these phenomena hinges on the conversion between charge and transverse spin currents -- the spin Hall effect (SHE) and its inverse (ISHE)~\cite{Kato2004,Wunderlich2005,Sinova2015}. In the SMR, both the SHE and ISHE play a concerted role, whereas in the SSE, thermal gradients across an interface result in a magnonic spin current, detected electrically by the ISHE. The SMR is an interface effect\cite{Isasa2014,Chen2013,JiaSTT2011}, whereas the SSE is explained as a bulk effect in which thermal magnons play an important role~\cite{Agrawal2014,Kehlberger2015,Rezende2016}. The SMR and SSE can be used to detect magnetic states, as recently demonstrated for a  frustrated magnet~\cite{Aqeel2015}.  
We performed a systematic study of the excitation and detection of spin currents in the spin spiral magnet CSO using the SMR and SSE in a transverse planar Hall-bar geometry. We investigated here two devices, S1 and S2, which consist of a platinum (Pt) strip deposited on a polished (111) surface of single crystal CSO (See Methods\ref{methods} for fabrication details). An optical image of the device is shown in Fig.~\ref{fig:2}a.

\begin{figure*}[Bpht]
\includegraphics[width=0.99\textwidth,natwidth=310,natheight=342]{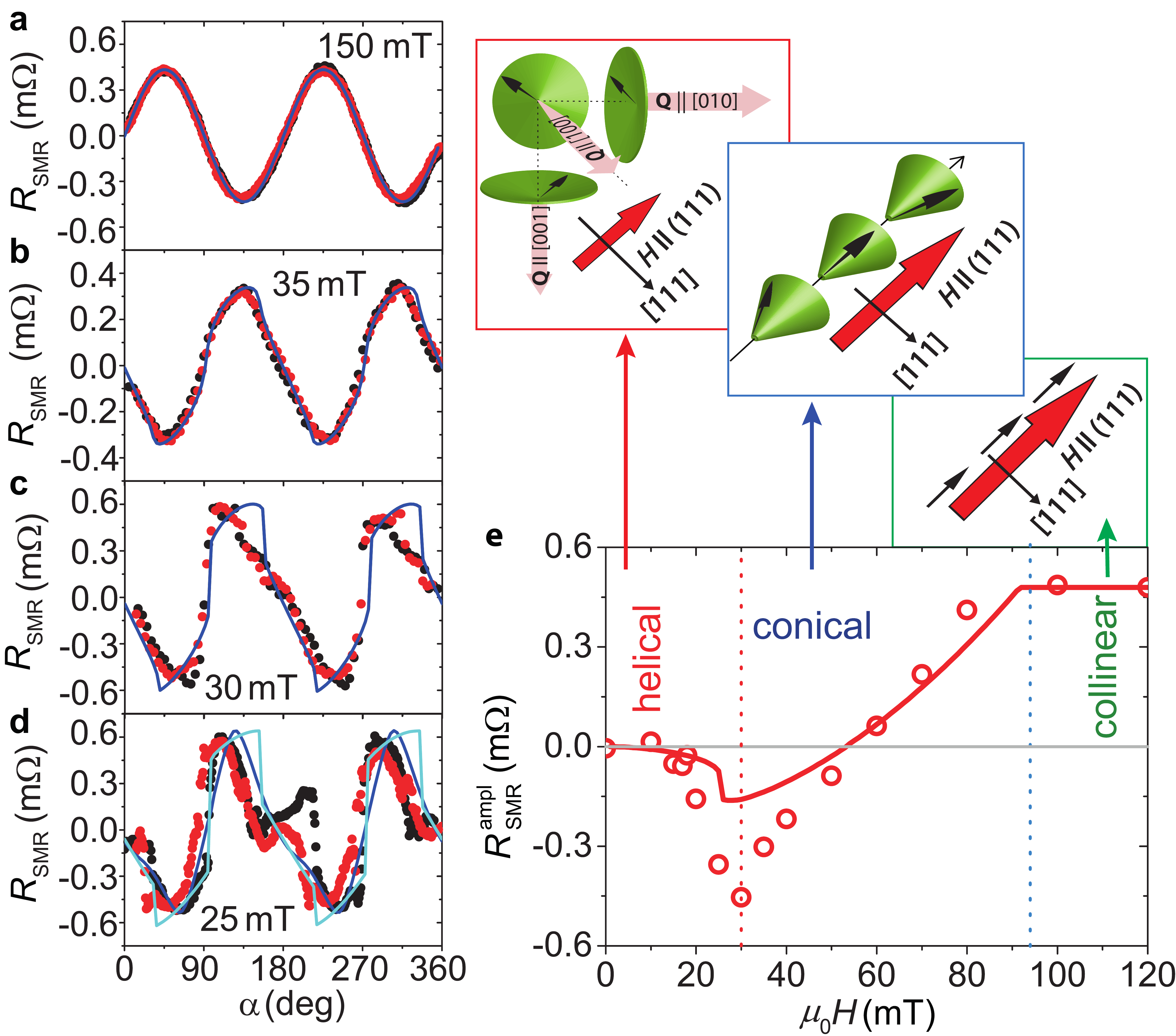}\caption{ \textbf{SMR in different magnetic states of CSO}.  \textbf{a-d}, Angular dependence of the SMR ($R_{\rm SMR}$) at 5~K in \textbf{a}, the collinear ferrimagnetic, \textbf{b,c}, conical  and \textbf{d}, helical magnetic states of CSO. The black and red data points show the trace and re-trace measurements, respectively, showing hysteresis at low applied magnetic fields. The solid curves fits to the data using the conical spiral Ansatz (equation~\ref{eq:conical}) (see \hyperref[suppl]{Supplementary Information} for details). The data are centred around zero and acquired for device S2. A more detailed evolution of the SMR signal can be found in Fig.S5 of \hyperref[suppl]{Supplementary}. \textbf{e}, Field dependence of the SMR signal, $R_{\rm SMR}^{\rm ampl}$ for the various magnetic orders which develop in CSO with increasing magnetic field. Here, $\bm Q$ represents the propagation wave vector and the magnetic field $\bm H$ is applied in the (111) plane parallel to the Pt$|$CSO interface. The transition between the different magnetic states of the CSO crystal is marked by vertical lines. The red line is the calculated amplitude of the SMR.}%
\label{fig:3}%
\end{figure*}

When a charge current $I$ is sent through a Pt strip, the SHE generates a transverse spin current. This spin current results a spin accumulation $\bm \mu_s$ at the Pt$|$CSO interface. This spin accumulation $\bm \mu_s$ can be absorbed or reflected at the interface depending on the direction of the magnetization $\bm M$ of CSO. When $\bm \mu_s \perp \bm M$, the electron spins arriving at the Pt$|$CSO interface partially absorbed  and when $\bm \mu_s \parallel \bm M$, spins will be fully reflected. The reflected spins generate an additional charge current by the ISHE. When $\bm M$ makes an angle with $\bm \mu_s$, this additionally generated charge current also has a component pointing in the transverse direction resulting in the transverse SMR response. Therefore, we expect to observe a dependence of the transverse Pt resistance on $\alpha$, the angle  between the applied current and the in-plane external magnetic field $\bm H$ that orients $\bm M$. We perform all transport measurements as function of $\alpha$, by rotating the device in a fixed field $\bm H$.  Because the change in the SMR voltage scales linearly with $I$, it can be detected by the first harmonic voltage response of the lock-in amplifier. Here, the SMR signal is shown after subtraction of an additional signal due to the ordinary Hall effect (see \hyperref[suppl]{Supplementary information} for details). The result of such a measurement is shown in Figs.~\ref{fig:2}b and \ref{fig:3}a. It clearly shows the $\sin(2\alpha)$ angular dependence expected for the collinear FM state~\cite{Chen2013}. We measured the angular dependence of the SMR at different temperatures in the FM state ($H>H_{c2}$) and observed a maximum SMR response around 5~K (see Fig.~\ref{fig:2}c). In order to explore the SMR response in different magnetic states of CSO, we set the temperature to 5~K and measured the angular dependence of the SMR for different external magnetic field strengths. 

The angular dependence of the SMR in the conical spiral state is close to the $\mathrm{\sin (2\alpha)}$ dependence observed at high fields and shows the presence of higher harmonics close to the transition to the helical state (see Figs.~\ref{fig:3}b,c). The most prominent feature is the change of the sign of the amplitude of the SMR, $R_{\rm SMR}^{\rm ampl}$ , in the conical spiral phase (cf. Fig.~\ref{fig:3}a with Figs.~\ref{fig:3}b,c): $R_{\rm SMR}^{\rm ampl}$ increases with field from a negative value at $H_{c1}$ to a positive value at $H_{c2}$. It changes sign at $\mu_0 H \sim 60$ mT   (see Fig.~\ref{fig:3}e). In the helical state, amplitude $R_{\rm SMR}^{\rm ampl}$ is negative: it equals zero in zero field and decreases with applied field. The angular dependence of the SMR in the helical state has sharp discontinuities and deviates strongly from the sinusoidal dependence (see Fig.~\ref{fig:3}d).  

This observed behavior can be understood as follows. For collinear magnets~\cite{Chen2013}, $R_{\rm SMR}~\propto~m_x~m_y$, where $m_x$ and $m_y$ are the in-plane components of the magnetization unit vector $\bm m$, parallel and orthogonal to the current direction, respectively. Since the spin relaxation length $\lambda~\sim~2$ nm in Pt is much smaller than the spiral period $\sim 50$ nm in CSO, this expression is valid {\em locally} for CSO.   The SMR for the non-collinear magnet is obtained by averaging $m_x m_y$ over the interface:
\begin{equation}
R_{\rm SMR} \propto \langle m_x m_y \rangle
\label{eq:RSMR}
\end{equation}
In the conical spiral phase
\begin{equation} \label{eq:conical}
\bm m = \cos \theta \bm e_3 + \sin \theta (\bm e_1 \cos \bm Q \cdot \bm x+\bm e_2 \sin \bm Q \cdot \bm x)
\end{equation} 
where $\bm e_1,\bm e_2$ and $\bm e_3$ are three mutually orthogonal unit vectors and $\theta$ is the cone angle. In the conical spiral state, both $\bm e_3$ and $\bm Q$ are parallel to the applied magnetic field, $\bm H$. Using equation~(\ref{eq:conical}), we obtain
\begin{equation}
 \langle m_x m_y\rangle = \frac{1}{4} \sin 2\alpha (3 \cos^2 \theta - 1)
\end{equation} 
which explains the (nearly) sinusoidal $\alpha$ dependence of SMR in the conical spiral state. In addition, $\cos \theta = \frac{H}{H_{c2}}$ (see \hyperref[suppl]{Supplementary Information}), so that $R_{\rm SMR}^{\rm ampl} \propto 3 \left( \frac{H}{H_{c2}}\right)^2-1$ increases with the magnetic field and changes sign at $H = \frac{H_{c2}}{\sqrt{3}}$, in good agreement with the experimental observations. The negative $R_{\rm SMR}$ close to $H_{c1}$ follows from the fact that for $\theta \sim 90^\circ$ spins in the conical spiral are nearly orthogonal to the magnetic field direction. Furthermore $R_{\rm SMR}$ remains constant for $H > H_{c2}$, as in the collinear phase $\theta = 0$, independent of the applied field strength.

The spin structure of the helical spiral state is more complex because of the deformation of the helix in an applied magnetic field and the presence of domains with different orientations of $\bm Q$. The angular and field dependence of the observed SMR can be qualitatively understood using equation~(\ref{eq:conical}), in which $\bm e_3$ and $\bm Q$ are not necessarily parallel to the field direction. Their orientations and the angle $\theta$ in each domain are found by numerical minimization of energy for a given applied field $H$ (see \hyperref[suppl]{Supplementary information}). We then added the contributions of the three domains. The result (blue line in Fig.~\ref{fig:3}), fits well the angular dependence of the observed SMR. Also the magnetic field dependence of the SMR in the helical state is qualitatively similar to our experimental observations: $R_{\rm SMR}=0$ at zero field due to cancellation of the contributions of the three domains (see \hyperref[suppl]{Supplementary Information}). It is negative at non-zero fields because $\theta$ is close to $90^\circ$ and it decreases  with the applied magnetic field because of the reorientation of  $\bm e_3$ and $\bm Q$.

\begin{figure*}[!htb]
\centering
\includegraphics[width=0.99\textwidth, clip]{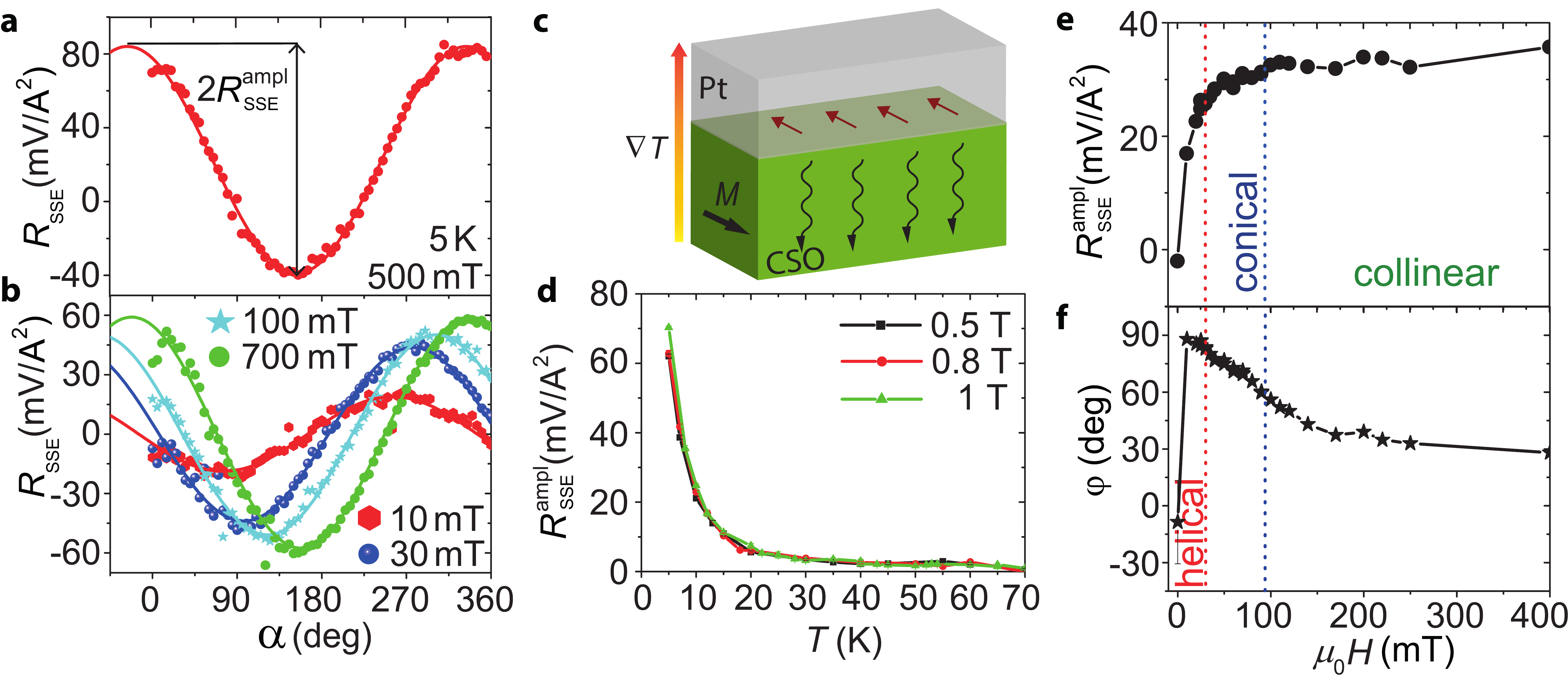}\caption{ \textbf{Electrical detection of the spin Seebeck effect (SSE)} \textbf{a,b},  Angular dependence of the SSE signal ($R_{\rm SSE}$) in the Pt$|$CSO. The solid lines show the  $\cos\left(\alpha+\varphi\right)$ fits. \textbf{c}, A schematic illustration of the SSE measured in the second harmonic response. The SSE is created by a current-induced thermal gradient $\nabla T$ across the Pt$|$CSO interface, generating magnons into the CSO crystal. The magnons create a spin accumulation close to the interface along the magnetization $\bm M$ of CSO. This spin accumulation is detected electrically by ISHE, resulting the SSE signal. \textbf{d}, Temperature dependence of  the spin Seebeck signal, $R_{\rm SSE}^{\rm ampl}$  for different applied magnetic field strengths. \textbf{e}, Amplitude $R_{\rm SSE}^{\rm ampl}$  and \textbf{f}, phase $\varphi$ of the SSE signal as a function of applied magnetic field at 5~K, respectively. The shown data are acquired for device S1 and for device S2 summarized in Fig.S6 of \hyperref[suppl]{Supplementary}. }%
\label{fig:4}%
\end{figure*}

In addition to the linear response of the SMR,  the SSE due to joule heating (for which $\Delta T \propto I^2$) also observed in the second harmonic response. In the SSE, thermally excited magnons spin polarize the electrons close to the interface, which is detected electrically by the ISHE (shown schematically in  Fig.~\ref{fig:4}c. As the generated spin accumulation is polarized along $\bm M$, the SSE/ISHE signal shows a  $\cos(\alpha)$ dependence by rotating $\bm M$, with  a full 360$^ \circ$ period.   The SSE signal is shown in Fig.~\ref{fig:4}a in which an additional non-zero phase $\varphi$ appears resulting in a $\cos\left(\alpha+\varphi\right)$ periodicity with an amplitude $R_{\rm SSE}^{\rm ampl}$. Both the amplitude $R_{\rm SSE}^{\rm ampl}$ and phase $\varphi$ vary with $H$ (see Fig.~\ref{fig:4}b).  The appearance of a non-zero $\varphi$ in the angular dependence of the SSE signal suggests that the bulk magnetization of CSO is not fully aligned along $\bm H$. When the magnetization is fully aligned along $\bm H$ (which ideally would be the case for the collinear ferrimagnetic state), we expect $\rm{\varphi= 0^\circ}$. Figure~\ref{fig:4}d shows that  $R_{\rm SSE}^{\rm ampl}$  increases with decreasing temperature and is observed to have maximum around 5~K (see \hyperref[suppl]{Supplementary information} for a plausible explanation). 
Figure~\ref{fig:4}e shows the field dependence of $R_{\rm SSE}^{\rm ampl}$ at 5~K. The amplitude $R_{\rm SSE}^{\rm ampl}$ is zero for $\mu_0 H=0~\rm T$ due to the absence of net magnetization in the helical spiral state.  $R_{\rm SSE}^{\rm ampl}$ grows much faster with the applied field in the helical phase than in the conical phase (see Fig.~\ref{fig:4}e), which may be attributed to the fact that the wave vector of the helical spiral has a component normal to the interface (along the [111] direction), resulting in cancellation  of the SSE signal sensitive to the in-plane component of the magnetization.  $R_{\rm SSE}^{\rm ampl}$ continues to grow with the field in the ferrimagnetic state, reaching the saturation at  ($\mu_0 H=4~\rm T$) and has the same sign as reported in literature~\cite{Aqeel2015} (see \hyperref[suppl]{Supplementary information} for details). Still, the weak field dependence of $R_{\rm SSE}^{\rm ampl}$ in the conical spiral phase is puzzling in view of the nearly linear dependence of the average magnetization on $H$.  Another puzzle is the field-dependence of the phase $\varphi$, which equals $\sim 90^\circ$ at low fields, decreases with increasing field and approaches $\sim 5^\circ$ in the field of 1~T (see Fig.~\ref{fig:4}f).

In conclusion,  we demonstrated that the SMR can be used for electric detection of transitions between complex spin states, such as the helical spiral and conical spiral phases of CSO. This technique was proved to be sensitive to the orientation of the spiral wave vector and to the magnitude of the cone angle. Our observation of the non-sinusoidal angular dependence of the SMR in the multi-domain helical spiral state and of the sign reversal of the SMR amplitude in the conical spiral state  provides new opportunities to conceive novel spintronic  devices based on this magnetoresistance effect. The observed complex angular and magnetic field dependence of the SMR is described remarkably well by our simple model of CSO.  The SSE also shows strong sensitivity to changes in magnetic ordering of CSO, although its origin remains unclear. It will be interesting to apply these techniques for detection of even more complex spin textures, such as skyrmions and merons. A technique with which one can observe nanosized objects by measuring electric currents would be indispensable for utilizing these topological defects as information carriers in next generation spintronics devices.

\section*{Methods}\label{methods} 
\renewcommand\refname{\vspace{-48pt}\setlength{\parskip}{12pt}}
\small {
\textbf{Fabrication}:
High quality CSO single crystals has been grown by a chemical vapor transport method~\cite{Belesi2010} with typical sizes 20-50mm$^3$. The crystal structure was characterized by a Bruker D8 Venture single crystal x-ray diffractometer. The magnetization of the crystals was measured  by a superconducting quantum interference device (SQUID) magnetometer. Before the device fabrication, the crystal surfaces were oriented along the (111) surface and polished to obtain the smallest surface roughness (see \hyperref[suppl]{Supplementary information}).  Two devices (S1 and S2) on two individually polished  (111) crystal surfaces (dimensions $\approx$ 4~mm*4~mm*2~mm) were prepared.  The Hall-bar device patterns were defined using three e-beam lithography steps, each followed by a standard deposition and lift-off procedures. The first step produces a Ti/Au~(5/40~nm) marker pattern, used to align the subsequent steps. The second step defines the platinum Hall-bar structure (5~nm), deposited by dc sputtering. The third step defines  Ti/Au~(5/80~nm) leads and bonding pads also deposited by d.c. sputtering in an Ar$^+$ plasma at an argon pressure of 3.3*10$^{−3}$~mbar and thickness in each step was measured by atomic force microscopy. 

\textbf{Measurements}:
All measurements were carried out in the transverse configuration as marked in Fig.~\ref{fig:2}a, by using two Stanford SR-830 lock-in amplifiers set at a reference  frequency of 17~Hz  (see \hyperref[suppl]{Supplementary information}). The lock-in amplifiers are used to measure the first and second harmonic signals. The same lock-in amplifiers are used to check third and fourth harmonic signals, at selected field and temperature regions. The current (ranged from 1 to 4~mA) was sent to the sample using a custom built current source and the response signal was recorded using a custom-built pre-amplifier (gain 10$^2$-10$^3$), before sending it back to the lock-in amplifier. All measurements were performed in a quantum design Physics Properties Measurement System (PPMS). The sample was rotated in the superconducting magnet of the PPMS with the stepper motor, such that the magnetic field varies in the plane of the sample as shown in Fig~\ref{fig:2}a.
}

\vspace{5mm}
\textbf{Acknowledgments}
\vspace{5mm}

\renewcommand\refname{\vspace{-48pt}\setlength{\parskip}{12pt}}
We would like to acknowledge J. Baas, H. Bonder, M. de Roosz and J. G. Holstein for technical assistance. This work is supported by the Foundation for Fundamental Research on Matter (FOM), NanoNextNL, a micro- and nanotechnology consortium of the government of the Netherlands and 130 partners, by NanoLab NL, InSpin EU-FP7-ICT Grant No 612759 and the Zernike Institute for Advanced Materials.

\vspace{5mm}
\textbf{Author contributions}
\vspace{5mm}

\renewcommand\refname{\vspace{-48pt}\setlength{\parskip}{12pt}}
A.A, B.J.v.W. and T.T.M.P. conceived the experiments. A.A. and N.V. designed the experiments. A.A. carried out the measurements. A.A. and N.V. elaborated the obtained data. The crystals were grown by A.A. and T.T.M.P. A.A. carried out the measurements.  A.A., B.J.v.W., M.M. and T.T.M.P. carried out the analysis. Theoretical modelling was developed by M.M. with simulations by M.M. and A.R. A.A. and M.M. wrote the paper, involving all co-authors.

\end{document}


~\label{suppl}

\title{Supplemental information: Electrical detection of spiral spin structures in Pt$|$Cu$_2$OSeO$_3$ heterostructures}
\author{A.\ \surname{Aqeel}}
\author{N.\ \surname{Vlietstra}}
\author{A.\ \surname{Roy}}
\author{M.\ \surname{Mostovoy}}
\author{B.\ J.\ \surname{van Wees}}
\author{T.\ T.\ M.\ \surname{Palstra}}

\keywords{}\maketitle

\renewcommand{\theequation}{S\arabic{equation}}
\renewcommand{\thefigure}{S\arabic{figure}}
\onecolumngrid
\section{Sample characteristics and measurement technique}~\label{suppl1}
In this section we provide additional information on the material properties and the  device fabrication of the Pt$|$CSO heterostructure. Firstly, the crystals were oriented by a single crystal x-ray diffractometer by focusing the x-ray beam on one corner. The Bruker Apex II software is used to rotate the goniometer such that the crystals were aligned along the [111] direction by using the orientation matrix obtained by collecting a dataset over a limited angular range. Some part of the crystal along the (111) plane was lapped away and then the (111) surfaces were polished in the following manner. 

\begin{figure*}[bhpt]
\centering
\includegraphics[width=0.8\textwidth, clip]{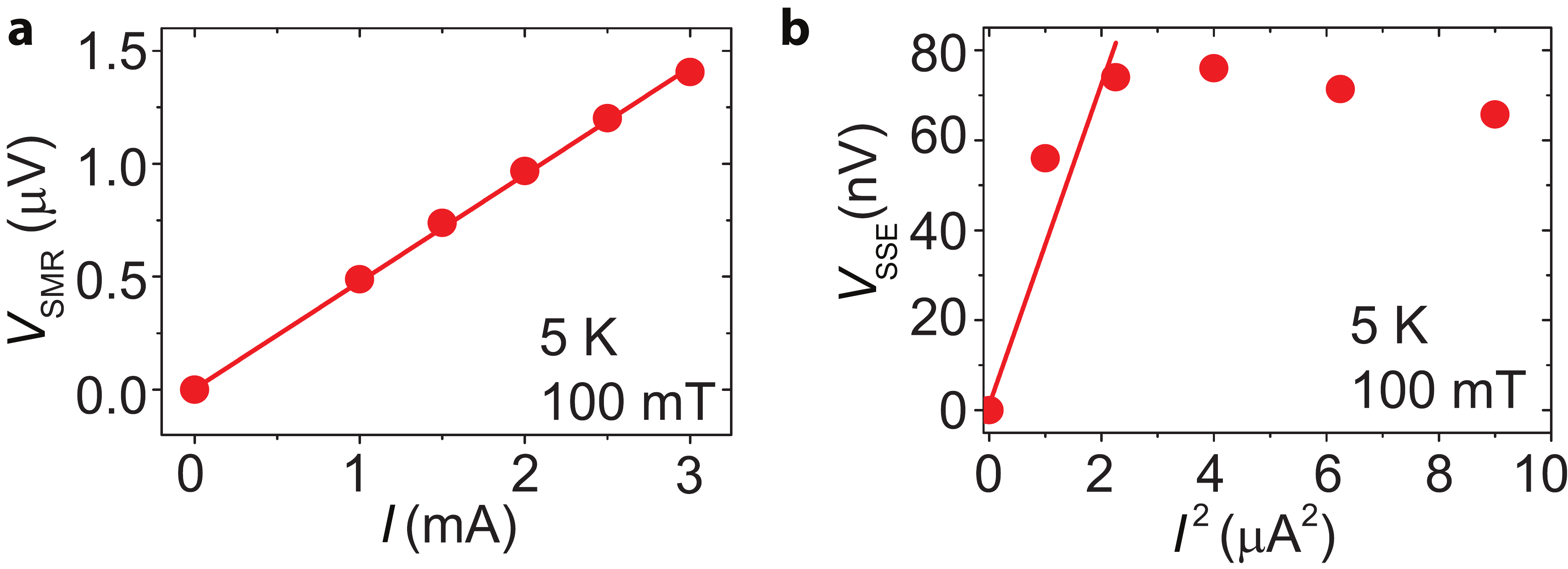}\caption{ \textbf{First and second harmonic signals in Pt$|$CSO} \textbf{a},  Current dependence of the first harmonic contribution $V_{\rm SMR}$ due to spin-Hall magnetoresistance, where $I$ is the ac current sent through the Pt Hall-bar.  \textbf{b}, Second harmonic signal $V_{\rm SSE}$ due to the spin Seebeck effect as a function of $I^2$, generated by current induced heating. Here, the solid lines indicate the linear fits.} 

\label{fig:5}%
\end{figure*}
\begin{figure*}[bhpt]
\centering
\includegraphics[width=0.99\textwidth, clip]{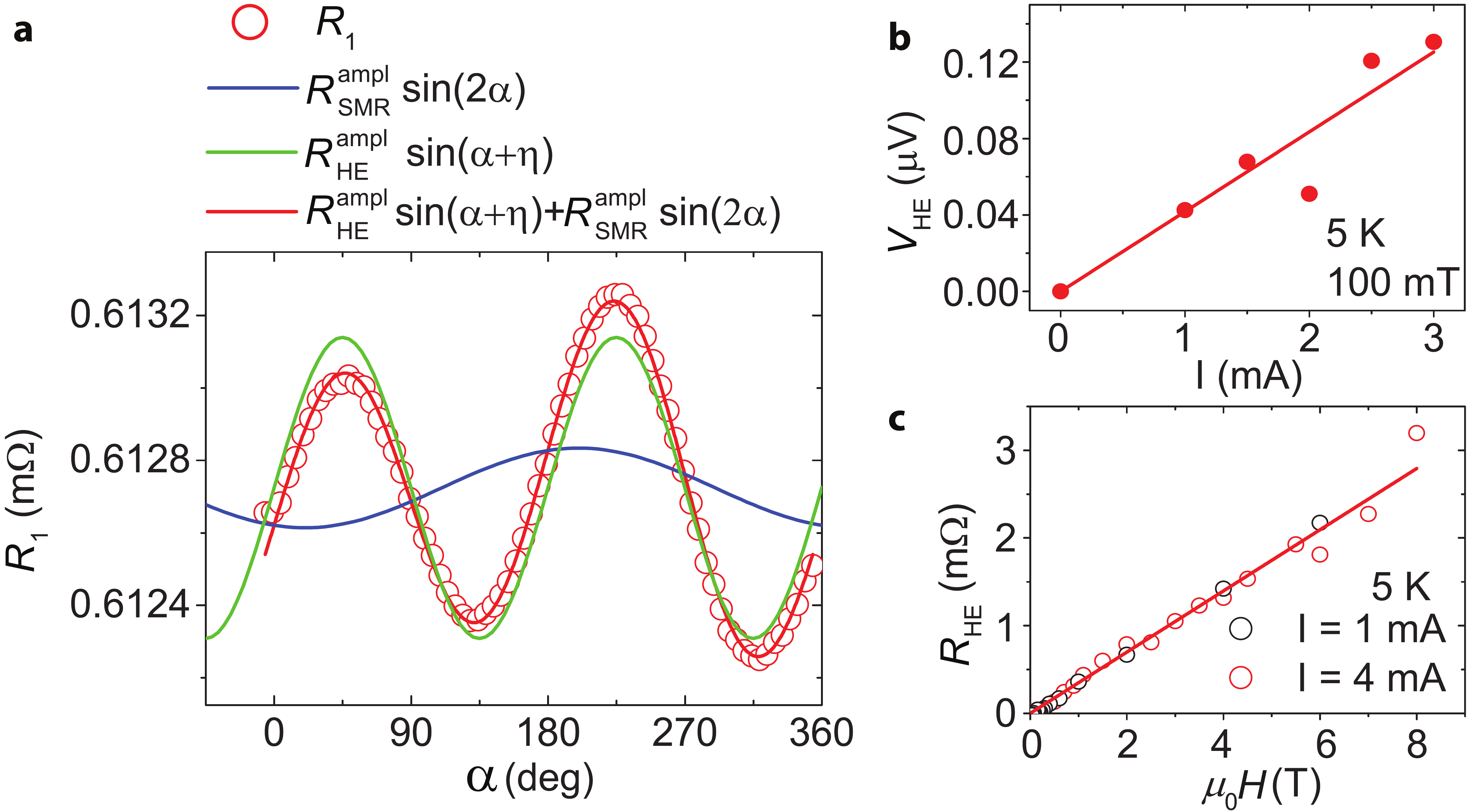}\caption{\textbf{SMR contribution in the first harmonic signal.} \textbf{a}, The angular dependence of the first order resistance response in the transverse configuration, $ R_1 = V_1/I$, for $I=4~\rm mA$ at 5~K in an
applied magnetic field of 400~mT. The $R_{\rm SMR}^{\rm ampl} \sin(2 \alpha)$ and $R_{\rm HE}^{\rm ampl} \sin(\alpha + \eta)$ curves illustrate the additive contributions from the expected SMR and  an additional ordinary Hall effect signals. \textbf{b}, The additional contribution in the first order resistance response $R_{\rm H}$ due to ordinary Hall effect  scales linearly with the applied current. \textbf{c},  $R_{\rm H}$, as expected for the Hall effect, scales linearly with the applied magnetic field. }%
\label{fig:6}%
\end{figure*}

The crystals were first slightly grinded with abrasive grinding papers (SiC P1200 - SiC P4000) by hand. After grinding, diamond particles were used with a sequence of 9~$\mu$m, 3~$\mu$m and 1~$\mu$m at 200~rpm (starting from slower speed of 100~rmp) for 15~mins, respectively. After each polishing step the crystals were cleaned with ethanol and acetone. As final polishing step, the surfaces were polished with colloidal silica OPS (oxide polishing suspension) with a particle size of 40~nm for 15~min. After polishing with silica particles, the crystals were quickly rinsed in water before drying. Then the crystals were cleaned by acetone and ethanol in an ultrasonics bath. We investigated two devices S1 and S2, fabricated in the same way but on two individually polished crystal surfaces.  

To measure the SMR and SSE simultaneously, we used a lock-in detection technique~\cite{Vlietstra2014}. By using this technique, we measured the SMR as first and the SSE as second order responses of the Pt$|$CSO system by sending an applied ac current ($I~\le$~4~mA) through the Pt Hall-bar. The output voltage signal can be  written as sum of first, second and higher order responses of $I$ as follows:
\begin{equation}
V(t)=R_{1}I(t)+R_{2}I^{2}(t)+R_{3}I^{3}(t)+R_{4}I^{4}(t)+\cdots, \label{eq1}%
\end{equation}
To measure the first and second order resistance  response, we measured the individual harmonic voltages by using lock-in amplifiers. When considering only first and second harmonic voltage signals, the first and second order resistance responses are defined as follows:  
\begin{equation}%
\begin{array}
[c]{c}%
R_{1}=\frac{V_{1}}{I_{0}}\\
R_{2}=\sqrt{2} \frac{V_{2}}{I_{0}^{2}}%
\end{array}
\text{ for }%
\begin{array}
[c]{c}%
\phi=0^{\text{o}}\\
\phi=-90^{\text{o}}%
\end{array}
\label{eqlock}%
\end{equation}
As the SSE is measured as second order resistance response, we defined here $V_2 = V_{\rm SSE}$ and $R_2 = R_{ \rm SSE}$. To check the contribution from the  higher harmonic responses, we also measured third and fourth harmonic signals at $I=4~\rm mA$ in different applied magnetic fields. We observed these signals to be negligible compared to the detected first and second harmonic signals. Therefore these higher harmonic signals do not have to be taken into account for the calculation of the first and second order response of the system~\cite{Vlietstra2014}.
In the linear response regime $I<2~\rm mA$, the SMR scales linearly and the SSE scales quadratically with the applied current as shown in Fig.~\ref{fig:5}a,b. However, at $I\geq 2~\rm mA$, the SSE no longer follows the expected quadratic trend. Nevertheless, we measured the SMR and SSE also by sending higher  currents till $I=4~\rm mA$. We observed a similar trend in the magnitude and phase change of the SSE at different currents (see Fig.~\ref{fig:8}g,h).

The angular dependence of the SMR and the SSE were studied by rotating an externally applied magnetic field in the
xy-plane of the CSO crystal~\cite{Schreier_APL_2013}.  The in-plane angle $\alpha$ of the magnetic field is defined relative to the applied current direction (x-axis) through the Pt Hall-bar, as indicated in Fig.~2a of the main text. All transport experiments were carried out by sending an ac-current ($I=2~\rm mA$) through Pt Hall-bar,

\section{First harmonic response in Pt on the CSO crystal}~\label{suppl2}

\begin{figure*}[Htbp]
\centering
\includegraphics[width=0.7\textwidth, clip]{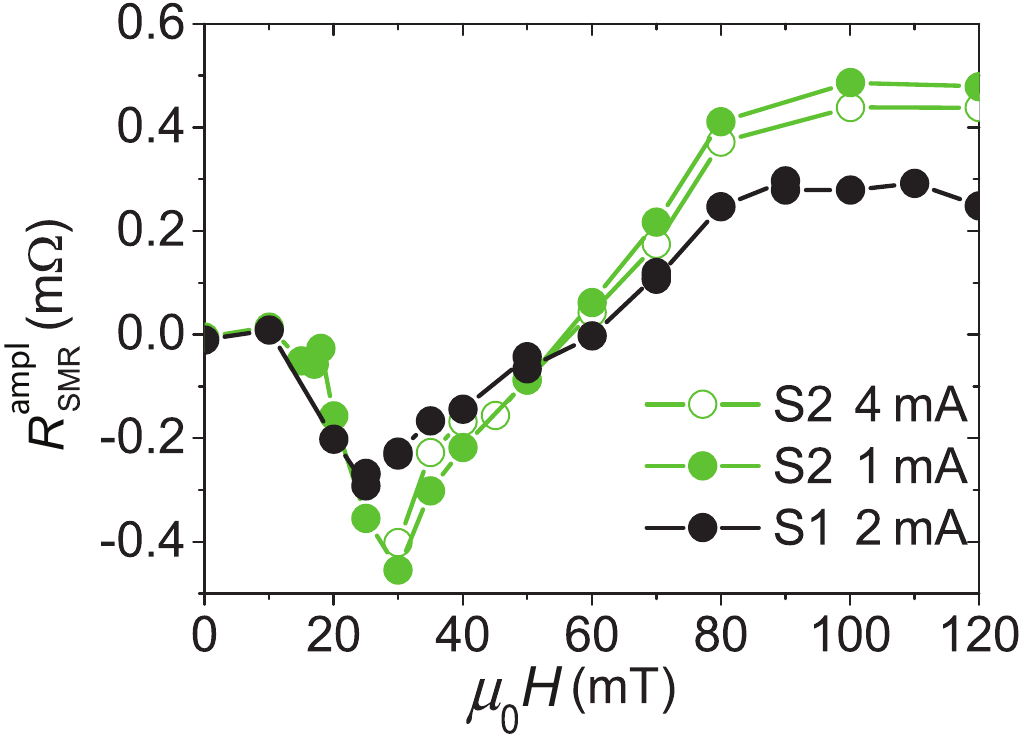}\caption{ \textbf{Comparison of the SMR signal amplitude in different devices.} The SMR signal as function of external magnetic field for devices S1 ($I= 2~\rm mA$) and S2 ($I = 1~\rm mA$, $\rm 4~mA$). }%
\label{fig:7b}%
\end{figure*}

\begin{figure*}[Htbp]
\centering
\includegraphics[width=0.99\textwidth, clip]{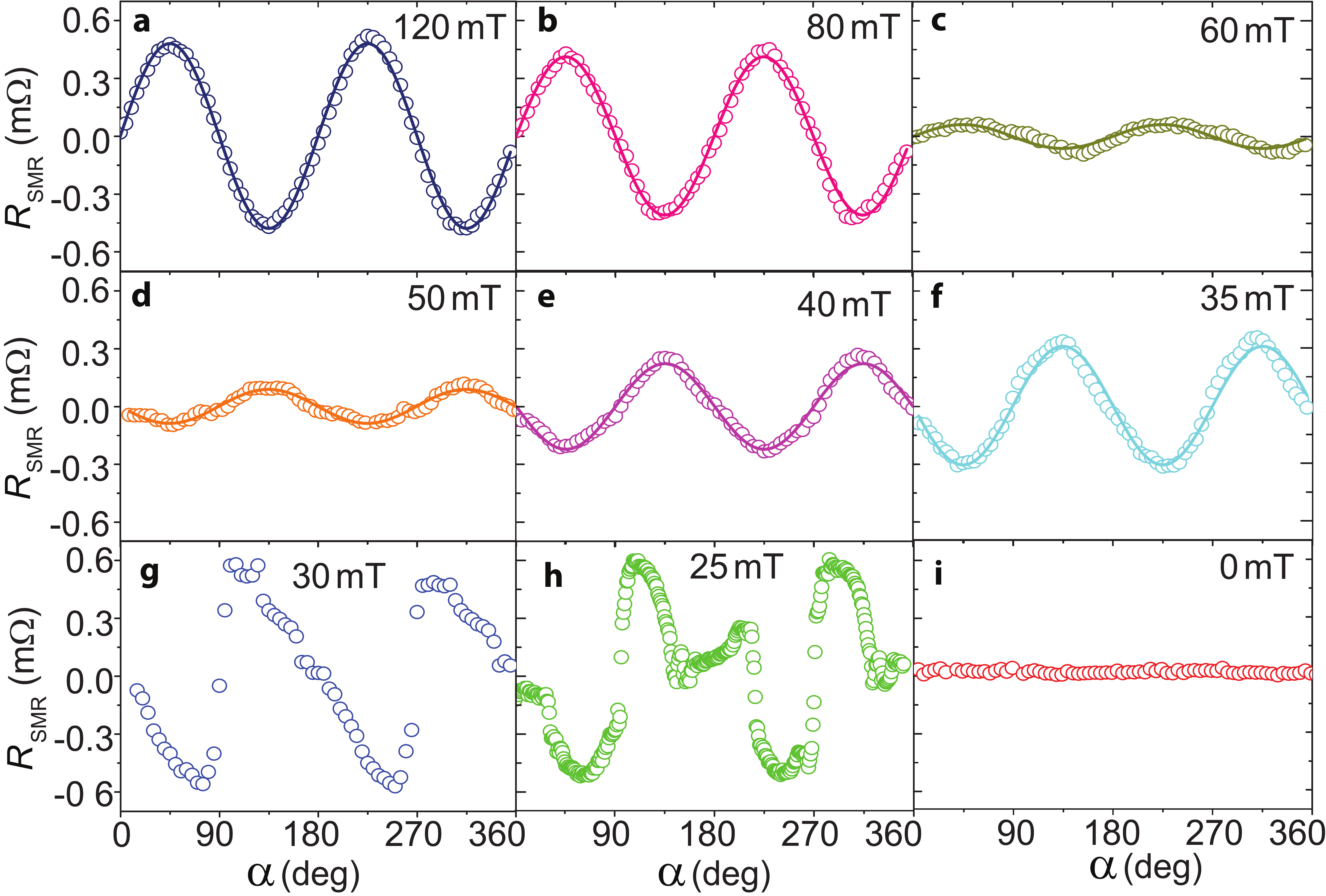}\caption{ \textbf{Evolution of the SMR signal in an applied magnetic field.} \textbf{a-i},  Change in the SMR signal $R_{\rm SMR}$ in Pt$|$CSO system by decreasing the applied magnetic field at current, $I=1~\rm mA$, through the Pt Hall-bar at 5~K. \textbf{a}, SMR in the ferrimagnetic state of CSO, \textbf{b-f}, in the conical magnetic state of CSO and \textbf{g-i}, in the helical magnetic state of CSO. The data shown here are centered around zero and acquired for device S2.}%
\label{fig:7}%
\end{figure*}

Here we discuss on the identification of different contributions in the first harmonic response and the method adopted to  separate the desired SMR contribution. We also discuss the influence of the applied magnetic field on the magnitude and the line shape of the SMR signal, as an addition to  the first part of the main text. The first order resistance response is determined  by measuring the first harmonic voltage transverse to the applied current direction on the Hall-bar structure, as described above. An example of such a measurement is shown by the red dots in Fig.~\ref{fig:6}a.

For the SMR the measured Pt resistance depends on the direction of average magnetization $M$  of CSO, as explained in the main text. Apart from the expected SMR, an additional signal  due to the ordinary Hall effect (HE) is obserevd in $R_1$, which is generated by a magnetic field component normal to the (111) plane of the CSO crystal due to a slight misalignment of the sample by an angle estimated for S1 and S2  to be $\sim 2^\circ$ and  $\sim 4^\circ$, respectively. The HE voltage has a $\sin(\alpha + \eta)$ angular dependence, where the phase $\eta$ is governed by the sample tilt direction. The ordinary Hall voltage of Pt$|$CSO scales linearly with the applied current and magnetic field, as expected (see Fig.~\ref{fig:6}b,c).

Our results for both devices S1 and S2 are reconcilable, despite the magnitude of the signals for device S2 is almost twice higher (see Fig.~\ref{fig:7b}), indicating that by optimizing the contact properties the signals could be enhanced further.  Figure~\ref{fig:7} shows the summarized data of the SMR as a function of angle $\alpha$ acquired for device S2 at 5~K. As explained in the main text, the SMR signal in the ferrimagnetic phase ($\mu_0H=120~\rm mT$)has a $\sin(2 \alpha)$ dependence (see Fig.~\ref{fig:7}b). When the field is reduced below the ferrimagnetic transition, the amplitude of the SMR signal decreases (see Fig.~\ref{fig:7}b,c) with the same sign. At $\mu_0H<60~\rm mT$, the SMR signal reverses its sign in the conical magnetic state of CSO (see Fig.~\ref{fig:7}d), a further decrease in magnetic field results in an increase of the amplitude of the SMR signal (see Fig.~\ref{fig:7}e,f). At the magnetic transition from conical to the helical state of CSO, the line shape of the SMR signal starts to deviate from a $\sin(2 \alpha)$ function (see Fig.~\ref{fig:7}f,g) and does not follow $\sin(2 \alpha)$ dependence in the helical phase (see Fig.~\ref{fig:7}h). The SMR signal fully disappear at $\mu_0H= 0~\rm T$ (see Fig.~\ref{fig:7}i).

\section{Continuum model for the SMR effect in Pt$|$CSO}~\label{suppl3}
We describe magnetic states of Cu$_2$OSeO$_3$ by the continuum model~\cite{Bak1980},
\begin{equation}~\label{S3}
\varepsilon = \frac{J}{2a} \sum_i \partial_i \bm m \cdot \partial_i \bm m + \frac{D}{a^2} \bm m \cdot \bm \nabla \times \bm m - \mu_0 \mu \bm m \cdot \bm H + \frac{K_1}{a^3} \sum_i m_i^4 + K_2 a \sum_i \partial_i^2 \bm m \cdot \partial_i^2 \bm m ,
\end{equation}
where the first term describes the FM exchange interaction, the second term is the Lifshitz invariant resulting from the Dzyaloshinskii-Moriya interaction and the third term is the Zeeman energy. The last two terms are the magnetic and spatial anisotropies allowed by the P$2_13$ symmetry of the crystal lattice, $a$ is the lattice constant and $\partial_i = \frac{\partial}{\partial r_i}$, $i  = x,y,z$.  

Substituting equation~(2) into the expression for the energy density equation~(\ref{S3}), we obtain
\begin{eqnarray}
\resizebox{0.91\textwidth}{!}{$\varepsilon = \left(\frac{JQ^2}{2a} + \frac{D}{a^2} \bm e_3 \cdot \bm Q\right) \sin^2 \theta -\mu_0 \mu \cos \theta \bm e_3 \cdot \bm H + \frac{K_1}{a^3} \left(A(\theta) + B(\theta) \sum_i e_{3i}^4\right) + \frac{1}{2}K_2 a \sin^2 \theta Q^2 \sum_i \left(Q^2 -   e_{3i}^2 Q_i^2 \right)$}
\end{eqnarray}
where $A( \theta ) = 3 \sin^2 \theta \left(\frac{1}{8} \sin^2 \theta +  \cos^2 \theta \right)$ and $B( \theta ) = \frac{3}{8} \sin^4 \theta - 3 \sin^2 \theta \cos^2 \theta + \cos^4 \theta$. We neglect the exchange anisotropy, $\frac{K_3}{a} \sum_i (\partial_i m_i )^2$, in   equation~(\ref{S3}), which reduces to $\frac{K_3}{2} \sin^2\theta \left( Q^2 - (\bm Q\cdot\bm e_{3})^2\right)$. Since $\bm e_3$ is (nearly) parallel to $\bm Q$ at all applied fields, this term has little effect on the magnetic state.

We first discuss the conical spiral state, in which $\bm e_3 \parallel \bm Q \parallel \bm H$. Neglecting the relatively small anisotropy terms, we obtain $Q a = \frac{D}{J}$ by minimizing $\varepsilon$ with respect to $Q$, while the minimization with respect to $\theta$ gives $\cos \theta = \frac{H}{H_{c2}}$, where $\mu_0 \mu H_{c2} = \frac{D^2}{J}$. 

The anisotropy terms are crucial for stabilization of the helical state. Due to the second term in equation~(S3), $\bm e_3$ and $\bm Q$ are nearly parallel in the helical state. Then both anisotropic terms are, essentially, equivalent to $K' a \sum_i Q_i^4$, which for $K' < 0$ favors the $[100], [010]$ and $[001]$ directions of the spiral wave vector. In an applied magnetic field the wave vector $\bm Q$, minimizing the energy of each domain, deviates from the corresponding crystallographic axis. $K'$ determines the critical field, $H_{c1}$, at which the transition from the helical to conical spiral state occurs.

The observed vanishing of the SMR signal in zero field is explained as follows. In the domain $\alpha$ ($\alpha = 1,2,3$), $\langle m_x m_y \rangle = \frac{1}{2} (\bm e_3^{(\alpha)} \cdot  \hat{x}) (\bm e_3^{(\alpha)} \cdot  \hat{y}) (3 \cos^2 \theta^{(\alpha)} - 1),$ where $\hat{x}$ and $\hat{y}$ are the unit vectors in the $x$ and $y$ directions. In zero field,  $\cos \theta^{(\alpha)} = 0$ in all domains and $e_3^{(1)} = (1,0,0)$,  $e_3^{(2)} = (0,1,0)$ and $e_3^{(3)} = (0,0,1)$. Adding the contributions of all three domains and assuming that they have the same volume, we obtain : $-\frac{1}{6} \sum_\alpha  (\bm e_3^{(\alpha)} \cdot  \hat{x}) (\bm e_3^{(\alpha)} \cdot  \hat{y}) =  -\frac{1}{6} (\hat{x} \cdot \hat{y}) = 0$.

Blue lines in figures~3a-d are the angular dependence of the SMR in different magnetic states of CSO obtained by numerical minimization of the energy~(S3) with respect to $\bm Q$ and $\bm e_3$, for $K_1 = 0$ and $K_2 = 0.07 J$. To reproduce the angular dependence of the multi-domain helical (Fig.~3d) we used two assumptions: (1) the three domains with different orientations of the spiral wave vector $\bm Q$ have the same volume (dark blue line) and (2) the domain with the lowest energy for a given $\bm H$ occupies the whole sample (light blue line).  Red line Fig.~3e shows the magnetic field dependence calculated for the same set of parameters. The model captures nicely all the essence of the data and correctly gives the value of the field, at which the amplitude of the SMR changes sign. The calculated amplitude of the SMR at low fields is, however, smaller than that observed in our experiment.
\begin{figure*}[Bpht]
\includegraphics[width=0.99\textwidth,natwidth=310,natheight=342]{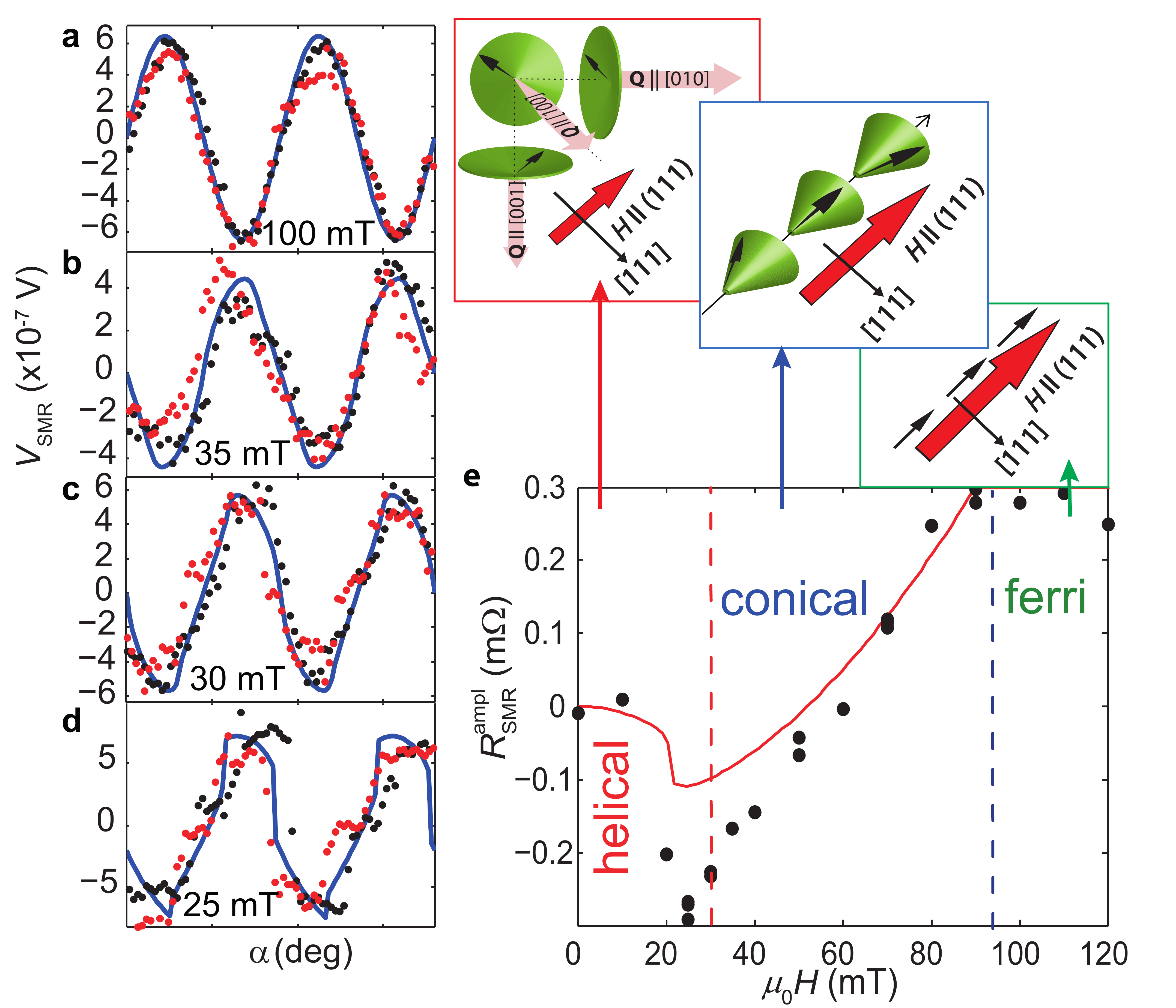}\caption{ \textbf{SMR in different magnetic states of CSO}.  \textbf{a-d}, Angular dependence of the SMR ($R_{\rm SMR}$) at 5~K \textbf{a}, in the collinear ferrimagnetic , \textbf{b}, conical  and \textbf{c,d}, helical  magnetic states of CSO. The black and red data points show the trace and re-trace measurements, respectively, showing hysteresis at low applied magnetic fields. The data are centred around zero and acquired for device S1. The solid curves are model fits to the data. \textbf{e}, Field dependence of the SMR signal, $R_{\rm SMR}^{\rm ampl}$ for the various magnetic orders which develop in CSO with increasing magnetic field. Here, $\bm Q$ represents the propagation wave vector and the magnetic field $\bm H$ is applied in the (111) plane parallel to the Pt$|$CSO interface. The transition between the different magnetic states of the CSO crystal is marked by vertical lines. The data acquired for device S2 are presented in Fig.~3 of the main text.}%
\label{fig:3ab}%
\end{figure*}

\section{Second harmonic response in Pt$|$CSO}~\label{suppl4}

\begin{figure*}[Htbp]
\centering
\includegraphics[width=0.99\textwidth, clip]{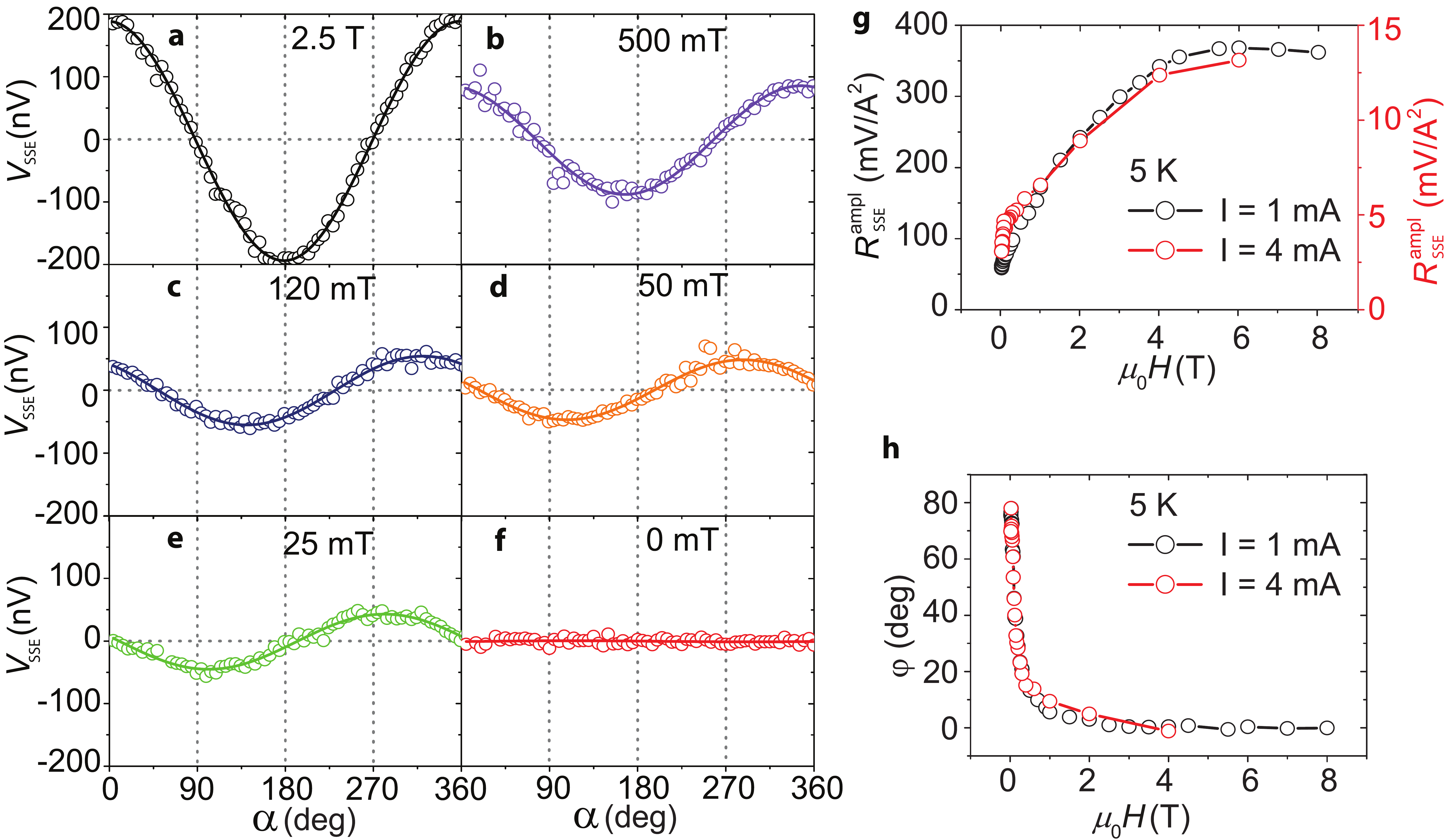}\caption{ \textbf{Evolution of the spin Seebeck signal with applied magnetic field.} \textbf{a-f}  Change in the magnitude and the line shape of the SSE response V$_{\rm SSE}$ in \textbf{a-c}, the ferrimagnetic , \textbf{d}, conical and \textbf{e,f}, in helical   magnetic states of  CSO.  \textbf{g}, Magnetic field dependence of the second harmonic response $R_{\rm SSE}^{\rm ampl}$ due to SSE by current induced heating in Pt$|$CSO.  \textbf{h}, Magnetic field dependence of  the phase $\varphi$ appearing in the angular dependence of V$_{\rm SSE}$ for different applied currents. The presented data are acquired from device S2 and centered around zero. }%
\label{fig:8}%
\end{figure*}
As shown in Fig.~4d in the main text, the SSE increases by decreasing the temperature below $T_c$, with a maximum signal observed at 5~K. 
The temperature dependence of the SSE resembles closely the temperature dependence observed in a frustrated magnetic system~\cite{Aqeel2015} and therefore can be explained in a similar way, by considering different sublattices. The associated  acoustic (ferromagnetic) and optical (antiferromagnetic) modes to different sublattices do not contribute to the SSE with the same sign and cancel to a large extent for temperatures close to $T_c$. Although by decreasing the temperature below $T_c$, the exchange splitting of the optical modes increases and they become increasingly depleted. The suppression of the thermal pumping of the optical modes in Pt$|$CSO thus leads to an apparent enhancement of the SSE at lower temperatures. This mechanism explains the low temperature sign change of the SSE of the ferrimagnetic insulator Gd$_3$Fe$_5$O$_{12}$ (GdIG)~\cite{Geprags2014}. It also accounts for the apparent suppression of the SSE in YIG at temperatures above 300 K~\cite{Uchida2014} and the enhancement in the SSE at low temperatures $T<T_c$ in a non-collinear magnetic insulator CoCr$_2$O$_4$~\cite{Aqeel2015}. A full theoretical modelling and interpretation is possible by considering the atomistic spin models although this is beyond the scope of this paper.

As shown in Fig.~4b,f in the main text for device S1 and in Fig.~\ref{fig:8} for device S2, the SSE detected as the second harmonic response changes phase as well as magnitude by decreasing the applied magnetic field (See Fig.~\ref{fig:8}) and at $\mu_0 H = 1~\rm mT$ the phase $\varphi$ is observed to be around 5$^\circ$. When the magnetic field is reduced further, the phase $\varphi$ significantly increases and reaches a value around 75$^\circ$ at the transition to the helical phase for $\mu_0 H = 30~\rm mT$ (see Fig.~\ref{fig:8}e,h). $V_{\rm SSE}$ fully disappears in a magnetic field of 0~T as shown in Fig.~\ref{fig:8}f. For both devices S1 and S2, a similar value of phase $\varphi$ is observed in the angular dependence of  $\mathrm{V_{SSE}}$ in different magnetic fields. Fig.~\ref{fig:8}g,h shows the amplitude and phase of the angular dependence of SSE, acquired from device S2 as function of field $H$ at $I=1~\rm mA$ (in the linear regime, where $V\rm_{SSE}$ scales linearly with $I^2$) and at $I=4~\rm mA$ (in the non-linear regime). For both current values, a similar trend in the amplitude $R_{\rm SSE}^{\rm ampl}$ and phase $\varphi$ is observed. The amplitude $R_{\rm SSE}^{\rm ampl}$ increases by increasing $H$ and starts to saturates around $\mu_0H=4~\rm T$ (see Fig.~\ref{fig:8}g). $R_{\rm SSE}^{\rm ampl}$ is more than four times larger at 4~T field than the signal observed at the conical to ferrimagnetic transition ($\mu_0 H = 94~\rm mT$). The phase $\varphi$ decreases by increasing the magnetic field and $\varphi \approx 0$ for $\mu_0H>1~\rm T$ (see Fig.~\ref{fig:8}h). It would be of great interest to develop a theoretical model and interpretation of the observed SSE signal however, this is outside the scope of this paper, where we provide a detailed summary of our experimental findings of the SSE in the Pt$|$CSO.
